\documentclass[12pt]{article}
\pdfoutput=1

\usepackage{amsmath}
\usepackage{amsfonts}
\usepackage{amscd}
\usepackage{amsthm}
\usepackage{setspace}
\usepackage{amssymb}

\usepackage{graphicx}
\usepackage{authblk}
\usepackage{caption}
\usepackage{ytableau}
\usepackage{mathtools}

\def\Z{\mathbb{Z}}

\def\P{\mathbb{P}}

\def\til{\tilde}

\begin{document}

\begin{titlepage}

\begin{flushright}
KEK-TH-2401
\end{flushright}

\vskip 1cm

\begin{center}

{\bf \Large Jacobian Calabi--Yau 3-fold and charge completeness in six-dimensional theory}

\vskip 1.2cm

Yusuke Kimura$^1$ 
\vskip 0.4cm
{\it $^1$KEK Theory Center, Institute of Particle and Nuclear Studies, KEK, \\ 1-1 Oho, Tsukuba, Ibaraki 305-0801, Japan}
\vskip 0.4cm
E-mail: kimurayu@post.kek.jp

\vskip 1.5cm
\abstract{We study aspects of an equivalent relation of the charge completeness in six-dimensional (6D) $\mathcal{N}=(1,0)$ supergravity theory and a standard assumption on the global structure of the gauge group involving F-theory geometry, recently proved by Morrison and Taylor. We constructed and analyzed a novel 6D supergravity theory, realized as F-theory, on an elliptically fibered Calabi--Yau 3-fold. Our construction yields a novel 6D theory with Mordell--Weil torsion $\mathbb{Z}_4\oplus\mathbb{Z}_4$. Furthermore, we deduce the gauge group and matter fields arising in the 6D F-theory model on the constructed elliptically fibered Calabi--Yau 3-fold. We also discuss the relations of the 6D F-theory model constructed in this study to stable degeneration and the dual heterotic string.
}

\end{center}
\end{titlepage}

\tableofcontents
\section{Introduction}
\par The swampland program attempts to determine the boundary that distinguishes string theories from those that appear to yield consistent quantum field theories at low energies but do not have a string theoretic realization. The latter theories belong to the swampland \cite{Vafa05, AMNV06, OV06} \footnote{Reviews of the swampland criteria can be found in \cite{BCV1711, Palti1903}.}. The swampland program has been a subject of vigorous research activities.  
\par Six-dimensional (6D) $\mathcal{N}=(1,0)$ supergravity theories possess a rich structure and are related to the swampland program. These theories have also been the subject of intensive research. The cancellation conditions of the gravitational, non-Abelian gauge and mixed anomalies impose strong constraints on the spectra of possible gauge groups and matter fields that can be formed in 6D $\mathcal{N}=(1,0)$ supergravity theories. Some 6D $\mathcal{N}=(1,0)$ supergravity theories have realizations in string theory through F-theory \cite{Vaf, MV1, MV2} \footnote{ Recent studies have investigated gauge groups formed in F-theory, for example \cite{Kumar200911, KMT1008, Kumar2010, MM2014, Morrison2016, Kimura1810, Cianci2018, Taylor2019, Kimura201902, Lee2019, Clingher2020, Karozas2020, Angelantonj2020, DelZotto2020, Klaewer2020, Morrison2021, Raghuram2021}.} on elliptically fibered Calabi--Yau 3-folds.
\par Related to the swampland and 6D $\mathcal{N}=(1,0)$ supergravity theories, very interesting aspects of charge completeness \footnote{The notion of the charge completeness was discussed by the authors in \cite{Banks2010}. Related conjectures concerning symmetries in quantum gravity were proved for theories in anti-de Sitter spaces with dual conformal field theories by the authors in \cite{Harlow2018}.} were studied in a recent study by Morrison and Taylor \cite{Morrison2021}. They showed in \cite{Morrison2021} that for 6D F-theory models, the property of charge completeness concerning the global structure of the gauge group is equivalent to a ``standard assumption'' in F-theory. F-theory is compactified in a space with a geometric structure referred to as an elliptic fibration. Fibers of the elliptic fibration are generically elliptic curves, and the modular parameter of the elliptic fibers is identified with axiodilaton in type IIB superstring theory. The ``standard assumption'' in F-theory \cite{AMrational, Aspinwall2000, MMTW} states \cite{Morrison2021} that the geometry of the elliptic fibration as a compactification space encodes the global structure of the gauge group. 
\par In some detail, when an elliptic fibration admits a structure referred to as a section, which is intuitively a copy of the base space of an elliptic fibration, one can ``add'' two sections. This addition operation equips the set of sections with a group structure known as the Mordell--Weil group. The Mordell--Weil group plays a role in the global structure of the gauge group according to the standard assumption in F-theory. 
\par The gauge algebra formed on the 7-branes is determined by the corresponding Kodaira singularity type \cite{Kod1, Kod2}, but the situation is more subtle for the global structure of the gauge group formed in the theory. Given an elliptic fibration with a global section, one can associate the Mordell--Weil group with that fibration. The structure of the Mordell--Weil group is a direct sum of finitely many copies of $\Z$ and the finite torsion part:
\begin{equation}
MW \cong \Z^n\oplus {\rm Tor}(MW).
\end{equation}
The gauge group formed in the 6D theory is denoted by $G$, and we use $G^0$ to denote the connected part of the gauge group $G$ to line up with the notations in \cite{Morrison2021}. We use $G_0$ to represent the simply connected non-Abelian part of $G$. Given a 6D $\mathcal{N}=(1,0)$ supergravity theory, which is realized as a 6D $\mathcal{N}=(1,0)$ F-theory on an elliptically fibered Calabi--Yau 3-fold and the elliptic fibration has the Mordell--Weil group $\cong$ $\Z^n\oplus {\rm Tor}(MW)$ (where ${\rm Tor}(MW)$ represents the Mordell--Weil torsion), the standard assumption in F-theory \cite{AMrational, Aspinwall2000, MMTW} states that the connected part $G^0$ is given as the quotient $G^0=(G_0\times U(1)^n)/H$, where $H$ satisfies the relation $H\cap G_0={\rm Tor}(MW)$, and ${\rm Tor}(\pi_1(G))$ coincides with the Mordell--Weil torsion ${\rm Tor}(MW)$. 
\par Morrison and Taylor formulated {\it the charge completeness hypothesis} in \cite{Morrison2021}, which states that given a minimally supersymmetric 6D or four-dimensional (4D) consistent theory of gravity, state exists for every possible value in the lattice of charges associated with the gauge group formed in that theory. They showed \cite{Morrison2021} that for 6D $\mathcal{N}=(1,0)$ F-theory realizations, the charge completeness hypothesis is equivalent to the standard assumption in F-theory, as stated previously. 
\par Morrison and Taylor \cite{Morrison2021} studied the standard assumption involving the Mordell--Weil group and an equivalent statement concerning the charge completeness for several families of 6D $\mathcal{N}=(1,0)$ supergravity theories, realized as F-theory on elliptically fibered Calabi--Yau 3-folds given by the Weierstrass equations with various Mordell--Weil torsions. Furthermore, they also explored stronger statements on massless fields that they confirmed hold for numerous 6D F-theory constructions \cite{Morrison2021}. 

\par Related to the framework \cite{Morrison2021} equating the standard assumption in F-theory on the global structure of the gauge group involving the Mordell--Weil group and a statement on charge completeness, {\bf the central goal of this note} is to discuss a construction of 6D $\mathcal{N}=(1,0)$ supergravity theories realized as F-theory on elliptically fibered Calabi--Yau 3-folds {\bf with a novel type of Mordell--Weil torsion} that were not explored in \cite{Morrison2021}. 

\vspace{5mm}

\par This paper has {\bf two themes}:
\begin{itemize}
\item[] 1. The first theme is to analyze 6D $\mathcal{N}=(1,0)$ supergravity theories realized as F-theory on elliptically fibered Calabi--Yau 3-folds {\bf with a new type of Mordell--Weil torsion} $\Z_4\oplus\Z_4$, which were not constructed in \cite{Morrison2021}. We studied these theories in relation to the charge completeness and the global structure of the gauge groups. 
\par Fibering Fermat quartic K3 surfaces, realized as a complete intersection in $\P^3\times \P^1$, over $\P^1$, and taking the Jacobian fibration yield such elliptically fibered Calabi--Yau 3-folds. We investigated 6D $\mathcal{N}=(1,0)$ F-theory on these Calabi--Yau 3-folds. 

\item[] 2. As the second theme, we will discuss the relation between 6D theories on elliptically fibered Calabi--Yau 3-folds and stable degeneration. The elliptic Calabi--Yau 3-folds constructed in this work admit K3 fibrations. Owing to this property, 6D F-theory on the Calabi--Yau 3-folds can be studied from the perspective of heterotic/F-theory duality \cite{Vaf, MV1, MV2, Sen, FMW}. As we will demonstrate, the Jacobian fibration of the Fermat quartic as K3 fibers of the Calabi--Yau 3-folds deforms into an elliptically fibered K3 surface that degenerates stably into a pair of rational elliptic surfaces. The Jacobian fibration of the Fermat quartic is obtained as a special limit of stable degeneration, wherein 7-branes become coincident. {\bf From the dual heterotic perspective, the coincident 7-branes on the F-theory side can be viewed as a nonperturbative effect of the insertion of 5-branes.} This viewpoint relates to the study of 6D $\mathcal{N}=(1,0)$ supergravity theories realized as F-theory on elliptically fibered Calabi--Yau 3-folds to dual heterotic theories. 

\end{itemize}

\par The main goal of this study is to construct an elliptically fibered Calabi--Yau 3-fold with Mordell--Weil torsion $\Z_4\oplus\Z_4$, and study 6D $\mathcal{N}=(1,0)$ F-theory on the 3-fold. The construction in this work yields a new 6D model because our construction provides a Calabi--Yau space with a novel type of Mordell--Weil torsion, which is a novelty of this work. Our construction does not belong to the families of models studied in \cite{Morrison2021}; therefore, our results expand the 6D theories studied in \cite{Morrison2021}.
\par We discuss the gauge algebra and matter fields formed in 6D $\mathcal{N}=(1,0)$ theory on the built elliptic Calabi--Yau 3-fold. 

\vspace{5mm}

\par We actually consider the complete intersection of two (2,1,1) hypersurfaces in the product of projective spaces $\P^3\times \P^1\times \P^1$ and then take the Jacobian fibration \footnote{A construction of the Jacobians of elliptic curves is discussed in \cite{Cas}.} to yield elliptically fibered Calabi--Yau 3-fold \footnote{Similar constructions of Calabi--Yau spaces can be found in \cite{Kimura1608, Kimura1907}.}. This approach enables one to determine the Mordell--Weil group of the Jacobian Calabi--Yau 3-fold. 
\par We chose $\P^3\times \P^1\times \P^1$ as the ambient space to ensure that the Calabi--Yau complete intersection, $Y$, of two (2,1,1) hypersurfaces in this ambient space admits not only an elliptic fibration, but also K3 fibration that is compatible with the elliptic fibration. We focus on the situation where K3 fibers are the Jacobian of the Fermat quartic. Then, the Jacobian Calabi--Yau 3-fold $J(Y)$ can also be viewed as a fibration of the Jacobian of the Fermat quartic over the base $\P^1$. Mathematical arguments utilizing modern techniques in algebraic geometry reveal that the Jacobian Calabi--Yau 3-fold $J(Y)$ has Mordell--Weil torsion $\Z_4\oplus\Z_4$ in this situation.
\par The approach that we utilize to analyze 6D $\mathcal{N}=(1,0)$ F-theory on the built Calabi--Yau 3-fold includes {\bf the three advantages}:
\begin{itemize}
\item[] 1. Taking the Jacobian of the (2,1,1) and (2,1,1) complete intersection in $\P^3\times \P^1\times \P^1$ with a special form of the defining equation enables the construction of elliptically fibered Calabi--Yau 3-fold with Mordell--Weil torsion $\Z_4\oplus\Z_4$, which provides a novel type of 6D $\mathcal{N}=(1,0)$ theory. The resulting 6D theory can be studied in the context of charge completeness, as discussed in \cite{Morrison2021}. 

\item[] 2. We analyzed the global geometric structure of the Calabi--Yau 3-folds by directly studying the defining equation. This approach does not rely on the standard Weierstrass techniques. This approach itself is novel, and can directly relate the analysis of gauge groups and matter fields to the global structure of the geometry of the compactification space. 

\item[] 3. By construction, the complete intersection of two (2,1,1) hypersurfaces in $\P^3\times \P^1\times \P^1$ admits a K3 fibration that is compatible with an elliptic fibration. Owing to this geometric property, 6D $\mathcal{N}=(1,0)$ theory on the Jacobians of the complete intersections can be analyzed from the dual heterotic perspective.

\end{itemize}

\vspace{5mm}

\par The non-Abelian gauge algebras formed in 6D $\mathcal{N}=(1,0)$ F-theory on an elliptically fibered Calabi--Yau 3-fold come from the 7-branes. These physical objects are wrapped on complex curves as components in a locus on the base complex surface, referred to as the discriminant locus \footnote{$\tau$ function of elliptic fibration in F-theory is identified with axiodilaton in the type IIB description. Elliptic fibers in the elliptic fibration degenerate along a complex codimension-one locus in the base complex surface, becoming singular fibers. Such locus is referred to as the discriminant locus, and 7-branes are wrapped on the components of this locus.}. The elliptic Calabi--Yau 3-fold develops Kodaira singularities \cite{Kod1, Kod2} along the discriminant locus. The Kodaira singularity in the elliptic fibration over a complex curve in the base complex surface corresponds to non-Abelian gauge algebra, in which the 7-branes wrapped on the complex curve support \cite{MV2, BIKMSV}. Table \ref{table singularity and gauge algebra} presents this correspondence. Monodromies can generate associated non-simply-laced algebras for some singularity types, as discussed in \cite{BIKMSV}. Therefore, to deduce the exact gauge algebra for such situations, an analysis of monodromy is also necessary \cite{Morrison2012}. 
\begingroup
\renewcommand{\arraystretch}{1.1}
\begin{table}[htb]
\begin{center}
  \begin{tabular}{|c|c|c|} \hline
Fiber type & Singularity & Non-Abelian gauge algebra \\ \hline
$I_n$ ($n\ge 2$) & $A_{n-1}$ & $\mathfrak{su}(n)$ or $\mathfrak{sp}(\lfloor \frac{n}{2} \rfloor)$ \\
$I_0^*$ & $D_4$ & $\mathfrak{so}(8)$, $\mathfrak{so}(7)$, or $\mathfrak{g}_2$. \\
$I^*_m$ ($m\ge 1$) & $D_{m+4}$ & $\mathfrak{so}(2m+8)$ or $\mathfrak{so}(2m+7)$ \\ 

$IV$ & $A_2$ & $\mathfrak{su}(3)$ or $\mathfrak{su}(2)$ \\
$III$ & $A_1$ & $\mathfrak{su}(2)$ \\ 

$II^*$ & $E_8$ & $\mathfrak{e}_8$ \\
$III^*$ & $E_7$ & $\mathfrak{e}_7$ \\
$IV^*$ & $E_6$ & $\mathfrak{e}_6$ or $\mathfrak{f}_4$ \\ \hline 
\end{tabular}
\caption{\label{table singularity and gauge algebra}Types of singular fibers, corresponding singularity types of the elliptic fibrations, and non-Abelian gauge algebras. We followed the notational convention used in \cite{Morrison2012}.}
\label{tabsingularity}
\end{center}
\end{table}
\endgroup
\par Complex curves as components in the discriminant locus, on which 7-branes are wrapped, generally intersect in the base surface, and at an intersection local enhancement of the singularity occurs. Consequently, a hypermultiplet arises at the intersection \footnote{Studies of matter representations arising from local enhancements of the singularities can be found, e.g. in \cite{BIKMSV, Katz1996, MTmatter}.}. The intersections of the 7-branes wrapped on the complex curves yield complex codimension-two loci on the base complex surface, and hypermultiplets arise at these points.
\par The Fermat quartic, $S_{Fermat}$, and the Jacobian fibration of the Fermat quartic, $J(S_{Fermat})$, are both K3 surfaces. The Jacobian Calabi--Yau 3-fold $J(Y)$ that we construct in this study has a K3 fibration whose K3 fiber is the Jacobian of the Fermat quartic $J(S_{Fermat})$. Except for the part to derive the defining equation of the Calabi--Yau 3-fold $J(Y)$, most of the properties of the Jacobian of the Fermat quartic $J(S_{Fermat})$ are not necessary for reading this note. Essentially, for physical application to the 6D model, we only utilize the property that the Jacobian of the Fermat quartic $J(S_{Fermat})$ admits an elliptic fibration with the Mordell--Weil group $\Z_4\oplus\Z_4$. We briefly summarize the geometric properties of the Jacobian of the Fermat quartic $J(S_{Fermat})$ in Appendix \ref{sectionA} as a reference for readers interested in the details of the derivation of the defining equation of the Calabi--Yau 3-fold $J(Y)$.

\par The construction of elliptically fibered Calabi--Yau 3-folds that we consider in this study extends to elliptic Calabi--Yau 4-folds. However, 4D $\mathcal{N}=1$ F-theory on elliptic Calabi--Yau 4-folds involves complications such as that D3-branes can also contribute to the gauge group formed in the theory, and quantum correction owing to a superpotential induced by the flux can lift the gauge group and matter fields \cite{Morrison2021}. We do not consider 4D theories in this note, and we focud on 6D $\mathcal{N}=(1,0)$ theories.

\vspace{5mm}

\par This paper is structured as follows: in section \ref{sec2}, we construct the Calabi--Yau 3-fold $J(Y)$ with the Mordell--Weil torsion $\Z_4\oplus\Z_4$, and we analyze physics of 6D $\mathcal{N}=(1,0)$ F-theory on the space. The main results are summarized in section \ref{subsec2.1}. Construction of the Calabi--Yau 3-fold $J(Y)$ with the Mordell--Weil torsion $\Z_4\oplus\Z_4$ is discussed in section \ref{subsec2.2.1}. The gauge algebra and the global structure of the gauge group formed in 6D F-theory on the constructed Calabi--Yau 3-fold are discussed in section \ref{subsec2.2.2}. The matter fields are deduced in section \ref{subsec2.2.3}. 
\par The elliptically fibered Calabi--Yau 3-fold $J(Y)$ has the base complex surface $\mathbb{F}_0=\P^1\times \P^1$. We will comment on 6D $\mathcal{N}=(1,0)$ F-theory on elliptic Calabi--Yau 3-folds over a general Hirzebruch surface $\mathbb{F}_n$ in section \ref{subsec2.3} by considering F-theory transition from $\mathbb{F}_0$ to $\mathbb{F}_n$ \cite{Seiberg1996, Witten1996} \footnote{The transition in 6D $\mathcal{N}=(1,0)$ F-theory on an elliptic Calabi--Yau 3-fold over a Hirzebruch surface can also be seen from the viewpoint of M-theory on ${\rm K3}\times S^1/\Z_2$ \cite{Seiberg1996, Witten1996}. In M-theory, the above transition in F-theory can be seen as a process in which small instantons are released as 5-branes from one of the two ends of $S^1/\Z_2$ and they undergo reabsorption into the other end \cite{Seiberg1996, Witten1996}.}.

\par Limit wherein an elliptically fibered K3 surface degenerates into two rational elliptic surfaces, intersecting along an elliptic curve, is known as the stable degeneration limit \cite{FMW, AM}. Heterotic/F-theory duality rigorously holds when K3 fibers take this limit on the F-theory side. A relation to the stable degeneration limit will be discussed in section \ref{sec3}. K3 fibers of the Calabi--Yau 3-fold $J(Y)$ with the Mordell--Weil group $\Z_4\oplus\Z_4$ are the Jacobian of the Fermat quartic, as we mentioned previously. Analogous to the situations discussed in \cite{Kimura1810, Kimura201902}, we will find that this K3 fiber is obtained as a limit of the stable degeneration at which 7-branes coincide. From the dual heterotic perspective, as discussed in \cite{Kimura1810, Kimura201902}, this can also be viewed as a non-perturbative effect owing to the presence of 5-branes. 
\par We state our concluding remarks in section \ref{sec4}.

\section{6D $\mathcal{N}=(1,0)$ F-theory models on Calabi--Yau 3-folds with Mordell--Weil torsion $\Z_4\oplus\Z_4$}
\label{sec2}
In this section, we construct an elliptically fibered Calabi--Yau 3-fold with a new type of Mordell--Weil torsion $\Z_4\oplus\Z_4$. We discuss a 6D F-theory on the resulting space, and relations to the charge completeness. Main results are summarized in section \ref{subsec2.1}. We discuss the construction of the elliptically fibered Calabi--Yau 3-fold, and we deduce the Mordell--Weil group in section \ref{subsec2.2.1}. We then determine the gauge algebra and matter spectra in sections \ref{subsec2.2.2} and \ref{subsec2.2.3}. 

\subsection{Summary of results}
\label{subsec2.1}
The main goal of this study is to construct an elliptically fibered Calabi--Yau 3-fold with Mordell--Weil torsion $\Z_4\oplus\Z_4$ and analyze the physics of 6D F-theory on the constructed Calabi--Yau space, including gauge algebra and matter fields. This construction adds a novel class to 6D $\mathcal{N}=(1,0)$ supergravity theories, realized as 6D F-theory with Mordell--Weil torsions analyzed in the context of charge completeness in \cite{Morrison2021}. The Calabi--Yau 3-folds built in this study also admit a K3 fibration that is compatible with elliptic fibration. Owing to this structure, 6D supergravity theories realized as 6D F-theory on the Calabi--Yau 3-folds in this study are also related to the stable degeneration limit. The aim of this study also includes an investigation of the relationship to the stable degeneration limit, as discussed in Section \ref{sec3}.

\par Now, we explain our strategy to construct an elliptic Calabi--Yau 3-fold. The construction process consists of two steps. First, to ensure that the space we build is Calabi--Yau, we construct a Calabi--Yau 3-fold, $Y$, by fibering the Fermat quartic $S_{Fermat}$ over $\P^1$. This is achieved by considering a complete intersection of two (2,1,1) hypersurfaces in the product $\P^3\times\P^1\times\P^1$, where we promote the equation of the Fermat quartic $S_{Fermat}$ to a Calabi--Yau 3-fold by replacing the coefficients depending on $\P^1$ with those depending on $\P^1 \times \P^1$. By construction, the Calabi--Yau 3-fold under discussion is genus-one fibered, where the genus-one fibration is given by natural projection onto the base $\P^1\times\P^1$. We provide the equation for the Calabi--Yau 3-fold $Y$ under discussion in Section \ref{subsec2.2.1}.
\par However, the resulting genus-one fibered Calabi--Yau space $Y$ does not have a global section; one can see this by arguing similar to that given in \cite{Kimura1603, Kimura1608}. Therefore, one cannot define the Mordell--Weil group of the Calabi--Yau $Y$. One needs to construct another Calabi--Yau space from $Y$ that has a global section.
\par To this end, as the next step, we consider the Jacobian fibration $J(Y)$ of the Calabi--Yau 3-fold $Y$. The central subject of this study is 6D $\mathcal{N}=(1,0)$ F-theory on this Jacobian Calabi--Yau 3-fold $J(Y)$. As demonstrated in Section \ref{subsec2.2.1}, the Jacobian Calabi--Yau 3-fold $J(Y)$ is described by the following double-cover description:
\begin{equation}
\label{double cover JCY in 2.1}
\boxed{
w^2=-t^2u^2 \lambda^4+\big(t^4u^4+1 \big) \lambda^2-t^2u^2.}
\end{equation}
Equation (\ref{double cover JCY in 2.1}) allows a transformation into the Weierstrass equation:
\begin{equation}
\label{Weierstrass JCY in 2.1}
\boxed{
y^2= \, x^3-\frac{1}{3}\big( t^8u^8+14\, t^4u^4+1 \big)\, x \, + \, \frac{2}{27}\big( t^4u^4+1 \big)\big( -t^8u^8+34\, t^4u^4-1 \big).
}
\end{equation}
By construction, the Jacobian Calabi--Yau 3-fold $J(Y)$ can be viewed as a fibration of the Jacobian of the Fermat quartic $J(S_{Fermat})$ (\ref{double cover JFermat in appB}) over $\P^1$. As previously stated in the Introduction, the Mordell--Weil group of the Jacobian Calabi--Yau 3-fold $J(Y)$ is isomorphic to that of the K3 fiber, that is, the Jacobian of the Fermat quartic $J(S_{Fermat})$, which is $\Z_4\oplus\Z_4$. We demonstrate this briefly in Section \ref{subsec2.2.1}. A genus-one fibration and Jacobian function have identical $\tau$ functions and discriminant loci. Therefore, the Calabi--Yau Jacobian fibration $J(Y)$ has a base $\P^1\times \P^1$, and the number of tensor multiplets formed in 6D F-theory on $J(Y)$ is one \cite{MV2}. $t$ and $u$ in the equations (\ref{double cover JCY in 2.1}) and (\ref{Weierstrass JCY in 2.1}) of the Jacobian Calabi--Yau 3-fold $J(Y)$ denote the inhomogeneous coordinates of the first $\P^1$ and the second $\P^1$ in the base $\P^1\times \P^1$, respectively.

\vspace{5mm}

\par Computing the discriminant, we deduce that the gauge algebra formed in 6D $\mathcal{N}=(1,0)$ F-theory on the Jacobian Calabi--Yau 3-fold $J(Y)$ is $\mathfrak{su}(4)^8$. This is discussed in Section \ref{subsec2.2.2}. 

\par Therefore, the intersections of 7-branes at which hypermultiplets arise in 6D F-theory have only $(A_3, A_3)$ collisions. Our analysis suggests that the adjoint representation ${\bf 15}$ arises at the collisions, as the anomaly free matter spectrum. This is discussed in Section \ref{subsec2.2.3}. 

\vspace{5mm}

\par The condition that an elliptic Calabi--Yau 3-fold has a K3 fibration compatible with the elliptic fibration only requires the base complex surface to be a Hirzebruch surface. However, when the base surface is a general Hirzebruch $\mathbb{F}_n$, which is not the product $\P^1\times\P^1$, the argument that the Mordell--Weil groups of the K3 fiber and the elliptic Calabi--Yau 3-fold are isomorphic does not necessarily hold. We comment on elliptic Calabi--Yau 3-folds over a general Hirzebruch surface as the base surface and 6D F-theory on these spaces in Section \ref{subsec2.3}. 

\vspace{5mm}

\par We stated that the Jacobian Calabi--Yau 3-fold $J(Y)$ that we consider in this study as compactification space for 6D $\mathcal{N}=(1,0)$ F-theory has the K3 fiber as the Jacobian fibration $J(S_{Fermat})$ of the Fermat quartic. By an argument similar to that given in \cite{Kimura1710, Kimura1810, Kimura201902}, one finds that the K3 fiber $J(S_{Fermat})$ corresponds to a special limit, which is a deformation of the sum of two extremal rational elliptic surfaces $X_{[4, 4, 2, 2]}$ intersecting along an elliptic curve $ X_{[4, 4, 2, 2]} \cup X_{[4, 4, 2, 2]}$. In line with the notation used in \cite{Kimura1710}, we use $X_{[4, 4, 2, 2]}$ to denote the extremal rational elliptic surface with two $I_4$ fibers and two $I_2$ fibers. Thus, the K3 fiber of the Calabi--Yau 3-fold $J(Y)$ is a deformation of stable degeneration. We find a special limit of the sum at which stacks of 7-branes meet yields the Jacobian K3 surface, $J(S_{Fermat})$. The singularity type of the K3 fiber becomes enhanced at this limit. From a heterotic viewpoint, this limit can be viewed as a non-perturbative effect owing to the presence of 5-branes. This is discussed in Section \ref{sec3}. 

\subsection{Calabi--Yau 3-fold with Mordell--Weil group $\Z_4\oplus\Z_4$ and 6D $\mathcal{N}=(1,0)$ F-theory}
\label{subsec2.2}

\subsubsection{Calabi--Yau 3-fold with Mordell--Weil group $\Z_4\oplus\Z_4$}
\label{subsec2.2.1}
Here, we construct an elliptically fibered Calabi--Yau 3-fold with Mordell--Weil group $\Z_4\oplus\Z_4$. The construction process consisted of two steps. 
\par First, we consider the complete intersection of two (2,1,1) hypersurfaces in $\P^3\times\P^1\times\P^1$ given by the following equation:
\begin{eqnarray}
\label{CY general in 2.2.1}
x_1^2+x_3^2+ 2tf(u)\, x_2x_4 = 0 \\ \nonumber
x_2^2+x_4^2+ 2tf(u)\, x_1x_3 = 0.
\end{eqnarray}
$[x_1:x_2:x_3:x_4]$ denotes the homogeneous coordinates of $\P^3$, $t$ denotes the inhomogeneous coordinate of the first $\P^1$ in $\P^3\times\P^1\times\P^1$, $u$ is an inhomogeneous coordinate of the second $\P^1$ in $\P^3\times\P^1\times\P^1$, and $f(u)$ is a degree-one homogeneous polynomial on the second $\P^1$ in $\P^3\times\P^1\times\P^1$. Mathematically, it is well known that the complete intersection of two hypersurfaces of degrees $(l_1, m_1, n_1)$ and $(l_2, m_2, n_2)$ in $\P^3\times\P^1\times\P^1$ is a Calabi--Yau 3-fold if and only if $(l_1, m_1, n_1)+(l_2, m_2, n_2)=(4,2,2)$. In particular, because $(2,1,1)+(2,1,1)=(4,2,2)$, the resulting space (\ref{CY general in 2.2.1}) is a Calabi--Yau 3-fold. Natural projection onto $\P^1\times\P^1$ yields a genus-one fibration. 
\par Utilizing an automorphism of the last $\P^1$ in $\P^3\times\P^1\times\P^1$, one has freedom to set $f(u)$ as $u$, and this does not have a physical consequence. It is well-known that one can choose three distinct points in $\P^1$ and send them to three distinct points in $\P^1$ under an appropriately chosen automorphism. $f(u)$ is of the form $u-a$, where $a$ is a point in $\P^1$. For this reason, one can send $a$ to zero under an appropriately chosen automorphism of $\P^1$. Therefore, the equation (\ref{CY general in 2.2.1}) can be rewritten as:
\begin{eqnarray}
\label{CY reduced in 2.2.1}
x_1^2+x_3^2+ 2tu\, x_2x_4 = 0 \\ \nonumber
x_2^2+x_4^2+ 2tu\, x_1x_3 = 0.
\end{eqnarray}
The projection onto the second $\P^1$ yields a K3 fibration compatible with the elliptic fibration, and the K3 fiber is the Fermat quartic (\ref{complete intersection Fermat in appB}). 

\par The next step, which completes the construction of an elliptically fibered Calabi--Yau 3-fold possessing a novel type of Mordell--Weil group, is to consider the Jacobian fibration, $J(Y)$, of the Calabi--Yau 3-fold $Y$ (\ref{CY reduced in 2.2.1}). Similar to the arguments in \cite{Kimura1603, Kimura1608}, one finds that the Calabi--Yau genus-one fibration $Y$ (\ref{CY reduced in 2.2.1}) lacks a global section, and the Mordell--Weil group is not defined. However, the Jacobian $J(Y)$ admits a global section, and the Mordell--Weil group is defined. 
\par A method to compute the Jacobian fibration of the complete intersection can be found in \cite{Kimura1603, Kimura1608, Kimura201905, Kimura201908}. By performing a simple computation by applying this method, one finds that the Jacobian Calabi--Yau 3-fold $J(Y)$ is given by equation (\ref{double cover JCY in 2.1}) \footnote{Generally, computing the Jacobian fibration of the complete intersection involves two steps: first, computing the associated double cover and then deducing the Jacobian fibration \cite{BM1401, MTsection} of the double cover. However, in the present situation, the associated double cover (\ref{double cover JCY in 2.1}) had a global section, since the coefficient of $\lambda^4$ is a perfect square \cite{MTsection}. Therefore, the double cover (\ref{double cover JCY in 2.1}) yields the Jacobian, and the second step is not required.}. By construction, the base complex surface is $\P^1\times\P^1$=$\mathbb{F}_0$ and a short computation shows that the discriminant is given as follows:
\begin{equation}
\label{discriminant in 2.2.1}
\boxed{
\Delta = 16\, u^4t^4 (ut-1)^4 (ut+1)^4 (ut-i)^4 (ut+i)^4.}
\end{equation}
After a computation similar to that given in \cite{Kimura1608}, one finds that the equation of the Jacobian Calabi--Yau 3-fold $J(Y)$ (\ref{double cover JCY in 2.1}) transforms into the Weierstrass equation (\ref{Weierstrass JCY in 2.1}).

\vspace{5mm}

\par Projection onto the last $\P^1$ in $\P^3\times\P^1\times\P^1$ induces a K3 fibration of the Jacobian Calabi--Yau 3-fold $J(Y)$. The K3 fibers of the K3 fibration of $J(Y)$ are the Jacobian of the Fermat quartic, $J(S_{Fermat})$ (\ref{double cover JFermat in appB}), which is evident from the definition of $J(Y)$ (\ref{double cover JCY in 2.1}). The elliptic fibration of the Jacobian K3 surface $J(S_{Fermat})$ has the Mordell--Weil group $\Z_4\oplus\Z_4$ \cite{Kimura1603} \footnote{This property also follows from the fact that the Jacobian of the Fermat quartic $J(S_{Fermat})$ is Shioda modular surface of level 4 \cite{Shiodamodular}. It was shown in \cite{Shiodamodular} that Shioda modular surface of level 4 has the Mordell--Weil group $\Z_4\oplus\Z_4$.}, as noted in the introduction. 
\par K3 fibers of the K3 fibration of the Jacobian Calabi--Yau 3-fold $J(Y)$ are the Jacobian of the Fermat quartic, $J(S_{Fermat})$, throughout the base curve $\P^1$. This implies that the complex structure of the K3 fibers is invariant over the base curve. It is expected that the Mordell--Weil group of the Jacobian Calabi--Yau 3-fold $J(Y)$ should be identical to that of the K3 fiber, whose complex structure is invariant over the base curve $\P^1$. One can demonstrate that this expectation is indeed true. Using a technique in algebraic geometry, we deduce that the Mordell--Weil groups of the elliptic Calabi--Yau 3-fold $J(Y)$ and the K3 fiber $J(S_{Fermat})$ are isomorphic from the argument as follows. Global sections of the Jacobian Calabi--Yau 3-fold $J(Y)$, when restricted to the K3 fiber, yield sections of the K3 fiber, which is the Jacobian of the Fermat quartic $J(S_{Fermat})$. This implies that the Mordell--Weil group of the Jacobian Calabi--Yau 3-fold $J(Y)$, $MW(J(Y))$, is a subgroup of the Mordell--Weil group of Jacobian of the Fermat quartic $J(S_{Fermat})$, which is $\Z_4\oplus\Z_4$. Namely, we have $ MW(J(Y))\subset MW(J(S_{Fermat}))=\Z_4\oplus\Z_4$. Therefore, to show that the two Mordell--Weil groups are indeed identical, it suffices to show that the Jacobian Calabi--Yau 3-fold $J(Y)$ has at least sixteen sections. 
\par One finds from the equation (\ref{double cover JCY in 2.1}) that the Jacobian Calabi--Yau 3-fold $J(Y)$ has at least sixteen sections. (\ref{double cover JCY in 2.1}) can be rewritten as $w^2+t^2u^2=\lambda^2\, [ -t^2u^2\lambda^2+\big(t^4u^4+1\big) ]$, and one finds from this that $\{w\pm i\,tu=0, \hspace{1mm} \lambda=0 \}$ yield two sections. By a symmetry argument, adding $t^2u^2\lambda^4$ to both sides of (\ref{double cover JCY in 2.1}), one finds that $J(Y)$ has additional two sections. Furthermore, since the right-hand side of (\ref{double cover JCY in 2.1}) splits into four linear factors, (\ref{double cover JCY in 2.1}) can also be rewritten as: $w^2=(tu-\lambda)(tu+\lambda)(tu\lambda-1)(tu\lambda+1)$. From this expression, one finds that $\{w=0, \hspace{1mm} tu-\lambda=0 \}$, $\{w=0, \hspace{1mm} tu+\lambda=0 \}$, $\{w=0, \hspace{1mm} tu\lambda-1=0 \}$, and $\{w=0, \hspace{1mm} tu\lambda+1=0 \}$ yield four sections. The equation (\ref{double cover JCY in 2.1}) can also be rewritten as $w^2=-t^2u^2(\lambda^2+1)^2+(t^2u^2+1)^2\lambda^2$. Then, from $w^2-(t^2u^2+1)^2\lambda^2=-t^2u^2(\lambda^2+1)^2$ one finds that $\{\lambda=i, \hspace{1mm} w=\pm (t^2u^2+1)i \}$ and $\{\lambda=-i, \hspace{1mm} w=\pm (t^2u^2+1)i \}$ yield four sections. In a similar fashion, by rewriting (\ref{double cover JCY in 2.1}) as $w^2-(t^2u^2-1)^2\lambda^2=-t^2u^2(\lambda^2-1)^2$, one deduces that $\{\lambda=1, \hspace{1mm} w=\pm (t^2u^2-1) \}$ and $\{\lambda=-1, \hspace{1mm} w=\pm (t^2u^2-1) \}$ yield four sections. From this argument, one learns that the Jacobian Calabi--Yau 3-fold $J(Y)$ has at least sixteen sections. Thus, we can conclude that the Mordell--Weil group of the Jacobian Calabi--Yau 3-fold (\ref{Weierstrass JCY in 2.1}) is $\Z_4\oplus\Z_4$. 
\par Most of the details of the Jacobian of the Fermat quartic $J(S_{Fermat})$ are not needed for the rest of this study. The details of the properties of the Fermat quartic $J(S_{Fermat})$ are provided in Appendix \ref{sectionA} for reference.

\subsubsection{Gauge algebra $\mathfrak{su}(4)^8$}
\label{subsec2.2.2}
From the discriminant (\ref{discriminant in 2.2.1}), we see that 7-branes in 6D $\mathcal{N}=(1,0)$ F-theory on the Calabi--Yau 3-fold $J(Y)$ are wrapped on the eight complex curves \footnote{It is straightforward to rewrite the Weierstrass equation (\ref{Weierstrass JCY in 2.1}) of the Jacobian Calabi--Yau 3-fold $J(Y)$ in a homogeneous form with homogeneous coordinates $[T:S]$ and $[U:V]$ for the first $\P^1$ and the second $\P^1$ in $\P^3\times\P^1\times\P^1$, respectively. (With these homogeneous coordinates, $t=\frac{T}{S}$ and $u=\frac{U}{V}$.) The discriminant (\ref{discriminant in 2.2.1}) can be rewritten, in the homogeneous form, as $\Delta=16\, S^4T^4U^4V^4 (TU-SV)^4 (TU+SV)^4 (TU-iSV)^4 (TU+iSV)^4$.}:
\begin{eqnarray}
\label{curves list in 2.2.2}
D_1 & = \{ t=0 \} \\ \nonumber
D_2 & = \{ t=\infty \} \\ \nonumber
D_3 & = \{ u=0 \} \\ \nonumber
D_4 & = \{ u=\infty \} \\ \nonumber
D_5 & = \{ tu-1=0 \} \\ \nonumber
D_6 & = \{ tu+1=0 \} \\ \nonumber
D_7 & = \{ tu-i=0 \} \\ \nonumber
D_8 & = \{ tu+i=0 \}.
\end{eqnarray}
The 7-branes wrapped on the curves $D_1, \ldots, D_8$ support the gauge algebra formed in the 6D F-theory. The singular fibers lying over the 7-branes wrapped on the eight curves $D_1$, $\ldots$, $D_8$ are of type $I_4$. From the Weierstrass equation (\ref{Weierstrass JCY in 2.1}), utilizing an argument similar to that given in \cite{Kimura1608}, it was found that all type $I_4$ fibers are split \cite{BIKMSV}. Therefore, we deduce that the non-Abelian gauge algebra formed on the 7-branes in the 6D F-theory is $\mathfrak{su}(4)^8$.
\par Because the Mordell--Weil group of the Calabi--Yau 3-fold $J(Y)$ is $\Z_4\oplus\Z_4$ as we saw in section \ref{subsec2.2.1}, the connected part of the 6D gauge group is the group quotient \footnote{The fundamental group of the connected part of the gauge group $G^0$, $\pi_1(G^0)$, is isomorphic to the Mordell--Weil torsion of the elliptic fibration for F-theory compactification, as discussed by the authors in \cite{AMrational}.}: $SU(4)^8/(\Z_4\oplus\Z_4$).
\par The types of the singular fibers lying over the complex curves $D_1, \ldots, D_8$, and the non-Abelian gauge algebras that the 7-branes wrapped on the curves support, are presented in Table \ref{table curves and gauge algebra in 2.2.2}.

\begingroup
\renewcommand{\arraystretch}{1.5}
\begin{table}[htb]
\begin{center}
  \begin{tabular}{|c|c|c|} \hline
Component & Fiber type & non-Abel. gauge algebra  \\ \hline
$D_{1,\ldots,8}$ & $I_4$ & $\mathfrak{su}(4)$  \\ \hline 
\end{tabular}
\caption{\label{table curves and gauge algebra in 2.2.2}The singular fiber types lying over the complex curves $D_1, \ldots, D_8$ on which 7-branes are wrapped, and the non-Abelian gauge algebra that the 7-branes wrapped on the curves support are listed.} 
\end{center}
\end{table}  
\endgroup

\vspace{5mm}

\par \noindent {\bf Remark :} We would like to remark that the gauge algebra formed in 6D F-theory on the Jacobian Calabi--Yau 3-fold $J(Y)$ does not contain a $\mathfrak{u}(1)$ factor coming from the Mordell--Weil group. The Jacobian Calabi--Yau 3-fold $J(Y)$ has the double-cover description (\ref{double cover JCY in 2.1}), and the coefficient of $\lambda^4$ is a perfect square. From a discussion given in \cite{MTsection}, one might at first expect that the Mordell--Weil rank would rise and $\mathfrak{u}(1)$ would form in the gauge algebra, but this is not the case for the model under discussion. Because the coefficient of $\lambda^3$ vanishes in (\ref{double cover JCY in 2.1}), $\mathfrak{su}(4)$ in fact forms, instead of $\mathfrak{u}(1)$, owing to a mechanism analogous to that discussed in section 3.1 of \cite{Kimura1907}. Therefore, the Mordell--Weil rank of the Jacobian Calabi--Yau 3-fold $J(Y)$ remains 0, and the gauge algebra acquires a $\mathfrak{su}(4)$ factor. The same argument applies to the constant term in the double-cover description (\ref{double cover JCY in 2.1}) by a symmetry argument. These are the origin of the two $\mathfrak{su}(4)$ factors in the gauge algebra.

\subsubsection{Matter hypermultiplets}
\label{subsec2.2.3}
Our analysis in Section \ref{subsec2.2.2} shows that 6D F-theory on the Calabi--Yau 3-fold $J(Y)$ has eight 7-brane stacks, which are wrapped on complex curves, $D_1, \ldots, D_8$ in the base $\P^1\times\P^1$. Each of the 7-brane stacks supports $\mathfrak{su}(4)$ algebra. The eight curves, $D_1, \ldots, D_8$, are all $\P^1$. 
\par Local enhancements of singularity types occur at the intersections of the pairs of the curves $D_1, \ldots, D_8$, giving rise to matter hypermultiplets at the intersections. Rewriting the equations of the curves (\ref{curves list in 2.2.2}) in homogeneous form, one finds after a short computation that the intersections occur at the following four points in the base $\P^1\times\P^1$:
\begin{eqnarray}
(0,0), & (0, \infty) \\ \nonumber
(\infty,0), & (\infty, \infty).
\end{eqnarray}
In fact, we have $(A_3, A_3)$ collisions at the intersections. By a symmetry argument, it is reasonable to expect that the matter hypermultiplets at $(A_3, A_3)$ collisions have identical representations. 
\par As we stated, the base complex surface of the Calabi--Yau 3-fold $J(Y)$ is $\P^1\times\P^1$, therefore, the number of tensor multiplets formed in the 6D theory is one, $T=1$. We use $V$ and $H$ to denote the number of vector multiplets and hypermultiplets in the 6D model, respectively, and the 6D anomaly cancellation condition \cite{GSW6d, Sagnotti, Erler, Sch6d} requires that $H-V=273-29=244$. As the gauge algebra formed on the 7-branes is $\mathfrak{su}(4)^8$, we have $V=120$. Thus, owing to the anomaly cancellation condition, the number of hypermultiplets is $H=273-29+120=364$. This condition is precisely satisfied, including neutral hypermultiplets and ${\bf 1}$, when six adjoint hypermultiplets ${\bf 15}$ of $\mathfrak{su}(4)$ arise at each of the four intersection points.
\par Therefore, we conclude that adjoint ${\bf 15}$ of $\mathfrak{su}(4)$ arises at the intersections of 7-brane stacks supporting $\mathfrak{su}(4)$ algebra.

\subsection{6D models on elliptic CY 3-folds over Hirzebruch surfaces $\mathbb{F}_n$ as base}
\label{subsec2.3}
We constructed an elliptic Calabi--Yau 3-fold $J(Y)$, whose base complex surface is $\P^1\times\P^1$. However, because an elliptic K3 surface is an elliptic fibration over $\P^1$ and K3 fibration over $\P^1$ yields a Calabi--Yau 3-fold, the condition that an elliptic Calabi--Yau 3-fold admits a compatible K3 fibration only requires that the base complex surface of the elliptically fibered Calabi--Yau 3-fold is a $\P^1$ bundle over $\P^1$. In other words, the condition of an elliptic Calabi--Yau 3-fold with a compatible K3 fibration does not limit the possible base surface to the direct product $\mathbb{F}_0=\P^1\times\P^1$; however, a general Hirzebruch surface $\mathbb{F}_n$ is also allowed \cite{MV1, MV2}. We comment on what happens when the base surface of an elliptic Calabi--Yau 3-fold is a general Hirzebruch surface $\mathbb{F}_n$. 
\par In 6D F-theory description, blowing up and blowing down points yield a transition from the base $\mathbb{F}_n$ to $\mathbb{F}_{n\pm 1}$ \cite{Seiberg1996}; therefore, via a sequence of blowing ups and blowing downs, a 6D F-theory on the base $\mathbb{F}_0=\P^1\times\P^1$ transitions to a 6D F-theory on the base $\mathbb{F}_n$ \footnote{Elliptic Calabi--Yau 3-fold over $\mathbb{F}_n$ does not exist for $n \textgreater 12$ \cite{Nak}.}.  
\par It seems plausible to expect that the process of blow-ups and blow-downs that transitions 6D F-theory on the base $\mathbb{F}_0=\P^1\times\P^1$ to a 6D theory on the base $\mathbb{F}_n$ does not affect the matter representations deduced for 6D F-theory on the base $\mathbb{F}_0=\P^1\times\P^1$ in section \ref{subsec2.2.3}, when intersections of 7-branes in the base occur well away from the points that are blown up or blown down. However, the Mordell--Weil group is likely to change when the base of the Calabi--Yau 3-fold transitions from $\mathbb{F}_0=\P^1\times\P^1$ to $\mathbb{F}_n$. 
\par Now, we discuss the above transition process quantitatively. The Jacobian Calabi--Yau 3-fold over the base $\P^1\times\P^1$, as constructed in Section \ref{subsec2.2.1}, has a natural generalization to an elliptic Calabi--Yau 3-fold over a general Hirzebruch surface $\mathbb{F}_n$, $J(\til{Y})$, as follows. Analogous to the construction in Section \ref{subsec2.2.1}, one can construct the Calabi--Yau complete intersection $\til{Y}$ in $\P^3\times \mathbb{F}_n$ given by:
\begin{eqnarray}
\til{f}_1\, x_1^2+\til{f}_1\, x_3^2+ 2\til{f}_2\, x_2x_4 = 0 \\ \nonumber
\til{f}_3\, x_2^2+\til{f}_3\, x_4^2+ 2\til{f}_4\, x_1x_3 = 0.
\end{eqnarray}
Here, $[x_1:x_2:x_3:x_4]$ denotes the homogeneous coordinates of $\P^3$ as before; $\til{f}_1$ and $\til{f}_2$ denote sections of line bundle $\mathcal{O}(C_1)$ over the base $\mathbb{F}_n$, and $\til{f}_3$ and $\til{f}_4$ denote sections of line bundle $\mathcal{O}(C_2)$ over the base $\mathbb{F}_n$, respectively, where divisors $C_1$ and $C_2$ are appropriately chosen so that the resulting space $\til{Y}$ is Calabi--Yau \footnote{To be more precise, they are subject to the constraint $C_1+C_2=-K_{\mathbb{F}_n}$, where $K_{\mathbb{F}_n}$ denotes the canonical divisor of the Hirzebruch surface $\mathbb{F}_n$.}. A computation analogous to that in section \ref{subsec2.2.1} shows that the Jacobian Calabi--Yau 3-fold $J(\til{Y})$ is given by:
\begin{equation}
\label{double cover JCY tilde in 2.3}
w^2=-\til{f_3}^2\til{f_4}^2 \lambda^4+\big(\til{f_1}^2\til{f_3}^2+\til{f_2}^2\til{f_4}^2 \big) \lambda^2-\til{f_1}^2\til{f_2}^2.
\end{equation}
The double-cover description (\ref{double cover JCY tilde in 2.3}) is transformed into the Weierstrass equation:
\begin{equation}
\label{Weierstrass JCY tilde in 2.3}
y^2= x^3-\frac{1}{3}\big( \til{f}_1^4 \til{f}_3^4+14\, \til{f}_1^2\til{f}_2^2\til{f}_3^2\til{f}_4^2+\til{f}_2^4 \til{f}_4^4 \big)\, x \,+ \, \frac{2}{27}\big( \til{f}_1^2\til{f}_3^2+\til{f}_2^2\til{f}_4^2 \big)\big( -\til{f}_1^4\til{f}_3^4+34\, \til{f}_1^2\til{f}_2^2\til{f}_3^2\til{f}_4^2-\til{f}_2^4\til{f}_4^4 \big).
\end{equation}
Because the Calabi--Yau 3-fold $J(\til{Y})$ (\ref{double cover JCY tilde in 2.3}) is a natural generalization of the Jacobian Calabi--Yau 3-fold $J(Y)$ (\ref{double cover JCY in 2.1}), it is plausible to assume that there is a transition from 6D F-theory on the Calabi--Yau 3-fold $J(Y)$ to 6D F-theory on $J(\til{Y})$ via a sequence of blow-ups and blow-downs.
\par We would like to note that the Weierstrass coefficients in the equation (\ref{Weierstrass JCY tilde in 2.3}) are indeed sections of line bundles $\mathcal{O}(-4K_{\mathbb{F}_n})$ and $\mathcal{O}(-6K_{\mathbb{F}_n})$, respectively, thus this confirms that (\ref{Weierstrass JCY tilde in 2.3}) correctly yields an elliptically fibered Calabi--Yau 3-fold over base $\mathbb{F}_n$. 

\vspace{5mm}

\noindent {\bf Non-Abelian gauge algebra $\mathfrak{su}(4)^2\oplus \mathfrak{su}(2)^4$:} The gauge algebra formed in 6D F-theory on the Calabi--Yau 3-fold $J(\til{Y})$ can be deduced, as we now demonstrate. It is straightforward to compute from the Weierstrass equation (\ref{Weierstrass JCY tilde in 2.3}) that the discriminant is given as:
\begin{equation}
\Delta = 16\, \til{f_1}^2\til{f_2}^2\til{f_3}^2\til{f_4}^2 \big(\til{f_1}\til{f_3}-\til{f_2}\til{f_4} \big)^4 \big(\til{f_1}\til{f_3}+\til{f_2}\til{f_4} \big)^4.
\end{equation}
Therefore, we find that the 7-branes are wrapped on six complex curves: $\til{D}_1= \{ \til{f_1}=0 \}$, $\til{D}_2= \{ \til{f_2}=0 \}$, $\til{D}_3= \{ \til{f_3}=0 \}$, $\til{D}_4= \{ \til{f_4}=0 \}$, $\til{D}_5= \{ \til{f_1}\til{f_3}-\til{f_2}\til{f_4} =0 \}$, $\til{D}_6= \{ \til{f_1}\til{f_3}+\til{f_2}\til{f_4}=0 \}$. Type $I_4$ fibers lie over the two components $\til{D}_5$ and $\til{D}_6$, and type $I_2$ fibers lie over the four components $\til{D}_1$, $\til{D}_2$, $\til{D}_3$ and $\til{D}_4$. Similar to the computation given in section \ref{subsec2.2.2}, type $I_4$ fibers are all split. Thus, we find that $\mathfrak{su}(4)^2\oplus \mathfrak{su}(2)^4$ gauge algebra forms on the 7-branes in the 6D theory. 

\vspace{5mm}

\noindent {\bf Matter representation:} $(A_3, A_3)$ and $(A_3, A_1)$ collisions occur at the intersections of the 7-brane stacks. Here, the base complex surface of the Calabi--Yau 3-fold $J(\til{Y})$ is a general Hirzebruch surface $\mathbb{F}_n$, $n\ne 0$, and not the product $\P^1\times\P^1$; this difference may affect the matter representations arising at the $(A_3, A_3)$ collisions, even though the collision type is identical to those studied in Section \ref{subsec2.2.3}. However, as stated previously, physical reasoning leads to a plausible expectation that matter representations would be left unaffected as long as the 7-branes intersect in the base complex surface well away from the points that are blown up or blown down. Whether matter representations identical to those deduced in Section \ref{subsec2.2.3} arise at the $(A_3, A_3)$ collisions in 6D F-theory on the Calabi--Yau 3-fold $J(\til{Y})$ is something that we would like to leave for future studies. 

\vspace{5mm}

\noindent {\bf Comment on Mordell--Weil group:} It seems difficult to compute the Mordell--Weil group of the Calabi--Yau 3-fold $J(\til{Y})$ (\ref{Weierstrass JCY tilde in 2.3}) constructed here, for the following reason. When the base surface is a Hirzebruch surface $\mathbb{F}_n$ which is not the product $\P^1\times\P^1$, then $\til{f}_i$, $i=2,4$, do not split into functions as $t\, f_i$, in sharp contrast to the situation in section \ref{subsec2.2}. As a result, the Jacobian $J(\til{Y})$ of the Calabi--Yau complete intersection $\til{Y}$ has a compatible K3 fibration; however, the complex structures of the K3 fibers are not invariant over the base $\P^1$. Owing to this complication for cases where the base surface is a general Hirzebruch surface $\mathbb{F}_n$ that is not the product of $\P^1$s, the argument presented in Section \ref{subsec2.2.1} does not necessarily apply. This complication prevents us from applying said to the elliptic Calabi--Yau 3-fold $J(\til{Y})$ over a general Hirzebruch surface as the base surface, which is not the product $\P^1\times\P^1$. Because the complex structures of the K3 fibers are not invariant over the base curve $\P^1$, we cannot relate the Mordell--Weil group of the K3 fiber to the Mordell--Weil group of the total Calabi--Yau 3-fold $J(\til{Y})$ by using an argument analogous to that in section \ref{subsec2.2.1}. 

\section{Relation to the stable degeneration limit and heterotic duals}
\label{sec3}
We demonstrate here that the Jacobian of the Fermat quartic, $J(S_{Fermat})$, as the K3 fiber of the Calabi--Yau 3-fold $J(Y)$ (\ref{Weierstrass JCY in 2.1}) constructed in Section \ref{subsec2.2} is obtained as a special limit of stable degeneration, wherein stacks of 7-branes meet and the singularity type is enhanced. This observation relates 6D $\mathcal{N}=(1,0)$ F-theory on the Calabi--Yau space $J(Y)$ to the stable degeneration limit and the heterotic dual, shedding some light on the relations among them. 
\par \noindent {\bf Relation to stable degeneration:} There is a K3 surface that stably degenerates into the sum of two rational elliptic surfaces $X_{[4, 4, 2, 2]}$ with two $I_4$ fibers and two $I_2$ fibers, intersecting along an elliptic curve $E$, $ X_{[4, 4, 2, 2]} \cup X_{[4, 4, 2, 2]}$ \cite{Kimura1710}. Here, we use the notation $X_{[4, 4, 2, 2]}$ to denote the extremal rational elliptic surface with two $I_4$ fibers and two $I_2$ fibers. A rational elliptic surface is said to be extremal when the elliptic fibration has a global section and the singularity type has a rank of eight. Such extremal rational elliptic surfaces were completely classified in \cite{MP}. The classification results in \cite{MP} are summarized in Appendix \ref{sectionB}. 
\par We consider a special limit of the sum $ X_{[4, 4, 2, 2]} \cup X_{[4, 4, 2, 2]}$, where two pairs of type $I_2$ fibers collide, and each pair of type $I_2$ fibers is enhanced to a type $I_4$ fiber. The sum $ X_{[4, 4, 2, 2]} \cup X_{[4, 4, 2, 2]}$ generally has four $I_4$ fibers and four $I_2$ fibers; however, in this special limit, the resulting elliptic K3 surface has six $I_4$ fibers. In fact, the properties of the resulting elliptically fibered K3 surface with a global section, and that it has six $I_4$ fibers, is sufficient to specify the complex structure; the complex structure of an elliptically fibered K3 surface with these properties is uniquely determined to be the Jacobian of the Fermat quartic $J(S_{Fermat})$ \cite{Kimura1603}, according to the classification result of the extremal K3 surfaces in \cite{SZ}. Briefly, the resulting elliptic K3 surface with six $I_4$ fibers, which was obtained as a special limit of the sum $ X_{[4, 4, 2, 2]} \cup X_{[4, 4, 2, 2]}$, is uniquely determined as the Jacobian of the Fermat quartic $J(S_{Fermat})$. From this argument, we conclude that the special limit yields the Jacobian of the Fermat quartic $J(S_{Fermat})$. In other words, the Jacobian of the Fermat quartic $J(S_{Fermat})$ can be viewed as a special limit of stable degeneration, at which an elliptic K3 surface stably degenerates into a pair of rational elliptic surfaces $X_{[4, 4, 2, 2]}$. Thus, because the K3 fiber of the Calabi--Yau 3-fold $J(Y)$ is the Jacobian of the Fermat quartic $J(S_{Fermat})$, 6D F-theory on the Calabi--Yau 3-fold $J(Y)$ is related to the stable degeneration limit. 

\newpage

\noindent {\bf Physical consequence that we can draw:} Because the intersection form on a rational elliptic surface is $U\oplus E_8$, the gauge algebra associated with the sum of two rational elliptic surfaces is the sum of the two algebras, $\mathfrak{a}_1$ and $\mathfrak{a}_2$, that are associated with two sublattices in $E_8$. Namely, $\mathfrak{a}_1, \mathfrak{a}_2\subset \mathfrak{e}_8$, and the gauge algebra associated with the sum of two rational elliptic surfaces is given as $\mathfrak{a}_1\oplus\mathfrak{a}_2 (\subset \mathfrak{e}_8\oplus\mathfrak{e}_8)$. Owing to this observation, when the K3 fiber stably degenerates over the base complex curve $\P^1$, the gauge algebra formed in 6D F-theory has a perturbative heterotic string interpretation. However, when the K3 fiber is the Jacobian of the Fermat quartic $J(S_{Fermat})$, which is a special limit of the stable degeneration as we just saw, the gauge algebra formed in 6D F-theory on the Calabi--Yau 3-fold $J(Y)$ is $\mathfrak{su}(4)^8$ as deduced in section \ref{subsec2.2.2}, which does not have a perturbative interpretation on the dual heterotic side. 
\par This strongly suggests that there is a nonperturbative effect present on the dual heterotic side, when the K3 fiber on the F-theory side becomes the Jacobian of the Fermat quartic $J(S_{Fermat})$ as a special limit of stable degeneration. We can draw a conclusion from an argument similar to those in \cite{Kimura1810, Kimura201902} as follows: {\bf the gauge algebra is enhanced when the K3 fiber on the F-theory side becomes the Jacobian of the Fermat quartic $J(S_{Fermat})$ as a consequence of the coincident 7-branes. When this is observed in the dual heterotic string theory, the enhancement in the gauge algebra is due to the nonperturbative effect of inserting 5-branes.} 

\vspace{5mm}
 
\par For eight-dimensional (8D) F-theory on K3 surface, elliptic fibers degenerate over finite points (to be more precise, twenty-four points counted with multiplicity) in the base $\P^1$, and 7-branes are located at these points. Collisions of singular fibers correspond to coincident 7-branes on the F-theory side, and from the dual heterotic viewpoint on two-torus $T^2$, coincident 7-branes correspond to the non-perturbative effect of the insertion of 5-branes \cite{MM2014}, as discussed in \cite{Kimura1810, Kimura201902}. 
\par An essentially identical argument applies to a 6D F-theory on an elliptic Calabi--Yau 3-fold without a qualitative alteration, and this fact follows from the following discussion. To begin with, let us start with an elliptically fibered Calabi--Yau 3-fold, that is obtained by fibering an elliptic K3 surface as the sum of copies of two rational elliptic surfaces intersecting along an elliptic curve, $X\cup X$, over $\P^1$. By construction, the complex structures of the K3 fibers are invariant over the base curve $\P^1$. The elliptic fiber of the K3 fiber becomes degenerate over some point in the base $\P^1$ of the K3 elliptic fibration. As the K3 fiber moves over the base curve $\P^1$ in the Calabi--Yau 3-fold, the trajectory of the point over which the elliptic fiber of the K3 surface has degenerated forms a complex curve, yielding a divisor in the base $\P^1\times\P^1$ of the Calabi--Yau 3-fold. 7-branes in the 6D F-theory on the Calabi--Yau space are wrapped on this complex curve. In this picture, collisions of singular fibers physically translate to 7-branes becoming coincident on the 6D F-theory side, analogous to the situations for 8D F-theory on an elliptic K3 surface. This argument particularly applies to a special limit of the stable degeneration at which pairs of type $I_2$ fibers collide in the K3 fiber, yielding the Jacobian of the Fermat quartic $J(S_{Fermat})$ that we previously discussed. Stacks of 7-branes overlayed with type $I_2$ fiber coincide on the 6D F-theory side on the Jacobian Calabi--Yau 3-fold $J(Y)$, at which point in the moduli space colliding $I_2$ fibers are enhanced to a $I_4$ fiber, and from the dual heterotic viewpoint this corresponds to the nonperturbative effect owing to the insertion of 5-branes. The theory on the heterotic side satisfies multiple 5-brane solutions.
\par {\bf The observation that we have gone through in this section provides another reason why the Calabi--Yau 3-fold $J(Y)$ is important in physical applications}, in addition to a 6D model with Mordell--Weil torsion $\Z_4\oplus\Z_4$ as we constructed in section \ref{subsec2.2}; {\bf application of the space $J(Y)$ casts a light on some aspects of the stable degeneration connecting 6D $\mathcal{N}=(1,0)$ F-theory models and the heterotic duals.}

\section{Concluding remarks}
\label{sec4}
We constructed an elliptically fibered Calabi--Yau 3-fold $J(Y)$ (\ref{Weierstrass JCY in 2.1}) with the Mordell--Weil group $\Z_4\oplus\Z_4$, and we discussed 6D $\mathcal{N}=(1,0)$ F-theory on the space in this study. The 6D model with Mordell--Weil torsion $\Z_4\oplus\Z_4$ provides a novel type of theory. We deduced that $\mathfrak{su}(4)^8$ gauge algebra forms in the 6D theory, and we also determined the global structure of the gauge group. We also deduced that adjoint hypermultiplet ${\bf 15}$ of $\mathfrak{su}(4)$ arises at the $(A_3, A_3)$ collisions. 
\par We constructed a novel type of elliptically fibered Calabi--Yau 3-fold, and hopefully it is clear at present that the approach that we utilized to analyze the geometry from the global perspective played some role in the construction.
\par The elliptic Calabi--Yau 3-fold constructed in this note has Mordell--Weil rank 0; therefore, the gauge algebra of the 6D F-theory on this space does not contain a $\mathfrak{u}(1)$ factor coming from the Mordell--Weil group. An investigation into a construction of elliptically fibered Calabi--Yau 3-fold with Mordell--Weil torsion $\Z_4\oplus\Z_4$ {\it with strictly positive Mordell--Weil rank}, and the 6D F-theory model on such space can be a direction of future study. If such space, i.e. Calabi--Yau 3-fold with Mordell--Weil torsion $\Z_4\oplus\Z_4$ {\it with strictly positive Mordell--Weil rank} does exist, the gauge algebra of 6D F-theory model on such space includes a $\mathfrak{u}(1)$ factor. 
\par A future study can focus on studying a deformation of the equation for the Calabi--Yau 3-fold $J(Y)$ that we constructed in this note to break one $\Z_4$ factor in the Mordell--Weil group. This can relate the 6D model with Mordell--Weil torsion $\Z_4\oplus\Z_4$ to the 6D models with Mordell--Weil torsion $\Z_4$ studied in \cite{Morrison2021}.

\section*{Acknowledgments}

We would like to thank Shigeru Mukai for discussions.

\appendix

\section{Jacobian of the Fermat quartic $J(S_{Fermat})$}
\label{sectionA}
We would like to briefly introduce the Jacobian of the Fermat quartic here. The Fermat quartic $S_{Fermat}$ is a K3 surface of Picard number 20, that is given as a degree-four hypersurface in $\P^3$ given by the following equation:
\begin{equation}
x_1^4+x_2^4+x_3^4+x_4^4=0 \subset \P^3.
\end{equation}
It is known that this K3 surface admits a genus-one fibration with six type $I_4$ fibers \cite{Kimura1603}. 
\par Aside from the description as a quartic hypersurface in $\P^3$, the Fermat quartic $S_{Fermat}$ has another presentation as the complete intersection of two bidegree (2,1) hypersurfaces in $\P^3\times \P^1$ of the following form \cite{Kimura1603}:
\begin{eqnarray}
\label{complete intersection Fermat in appB}
x_1^2+x_3^2+ 2t\, x_2x_4 = 0 \\ \nonumber
x_2^2+x_4^2+ 2t\, x_1x_3 = 0.
\end{eqnarray}
Here, $[x_1:x_2:x_3:x_4]$ denotes the homogeneous coordinates of $\P^3$, and $t$ yields the inhomogeneous coordinate of $\P^1$. Projection onto $\P^1$ yields a genus-one fibration $f: S_{Fermat}\rightarrow \P^1$. This presentation is useful when we construct Calabi--Yau 3-folds whose K3 fiber is the Jacobian of the Fermat quartic as in section \ref{subsec2.2}. 
\par The Jacobian fibration, $J(f): J(S_{Fermat})\rightarrow \P^1$, yields the Jacobian of the Fermat quartic, $J(S_{Fermat})$ with an elliptic fibration. The Jacobian fibration, $J(f)$, of the genus-one fibration $f$ admits a global section possessing six type $I_4$ fibers \cite{Kimura1603}. Equation of the Jacobian fibration $J(f)$ of the Jacobian of the Fermat quartic, $J(S_{Fermat})$, is given as follows \cite{Kimura1603}:
\begin{equation}
\label{double cover JFermat in appB}
w^2=-t^2\,\lambda^4+(t^4+1)\, \lambda^2-t^2,
\end{equation}
which is a double cover of $\P^1\times\P^1$ branched along a (4,4) curve. The equation (\ref{double cover JFermat in appB}) transforms into the Weierstrass equation as \cite{Kimura1603}:
\begin{equation}
\label{Weierstrass JFermat in appB}
y^2=\frac{x^3}{4}-\frac{(t^8+14t^4+1)x}{12}+\frac{t^{12}-33t^8-33t^4+1}{54}.
\end{equation}
The elliptic fibration (\ref{Weierstrass JFermat in appB}) has the Mordell--Weil group $\Z_4\oplus\Z_4$ \cite{Kimura1603} owing to the classification result of the extremal fibrations of K3 surfaces \cite{SZ}. Because the Jacobian of the Fermat quartic $J(S_{Fermat})$ (\ref{double cover JFermat in appB}) is Shioda modular surface of level 4 \cite{Shiodamodular} \footnote{Discussion of aspects of Shioda modular surfaces can be found in the first chapter of \cite{Mumford}.}, this also follows from a result derived in \cite{Shiodamodular}. 
\par We fibered the Jacobian $J(S_{Fermat})$ of the Fermat quartic, (\ref{double cover JFermat in appB}), over $\P^1$ to build an elliptically fibered Calabi--Yau 3-fold. As we demonstrated in section \ref{subsec2.2.1}, the resulting Calabi--Yau 3-fold has the Mordell--Weil group $\Z_4\oplus\Z_4$, that is isomorphic to the Mordell--Weil group of the Jacobian fibration $J(S_{Fermat})$ (\ref{double cover JFermat in appB}).

\section{Classification of extremal rational elliptic surfaces}
\label{sectionB}
The types of the singular fibers of the extremal rational elliptic surfaces were classified \cite{MP}. These are presented in Table \ref{classificationRES in appB}. The singular fiber types uniquely determine the complex structures of extremal rational elliptic surfaces, except those with two type $I_0^*$ fibers \cite{MP}. The complex structures of such rational elliptic surfaces possessing two type $I_0^*$ fibers depend on the j-invariants of the elliptic fibers, that are constant over the base $\P^1$ for the surfaces with two type $I_0^*$ fibers \cite{MP}. In the notation used in the column at the extreme left of Table \ref{classificationRES in appB}, we simply utilized $n$ to represent a type $I_n$ fiber, and $m^*$ to denote a $I^*_m$ fiber. Our notational convention followed those used in \cite{Kimura1710, Kimura1810}.

\begingroup
\renewcommand{\arraystretch}{1.5}
\begin{table}[htb]
\centering
  \begin{tabular}{|c|c|c|} \hline
$
\begin{array}{c}
\mbox{Extremal rational}\\
\mbox{elliptic surface} 
\end{array}
$ & Singular fiber type & Singularity type \\ \hline
$X_{[II, \, II^*]}$ & $II$, $II^*$ & $E_8$  \\ \hline
$X_{[III, \, III^*]}$ & $III$, $III^*$ & $E_7A_1$  \\ \hline
$X_{[IV, \, IV^*]}$ & $IV$, $IV^*$ & $E_6A_2$ \\ \hline
$X_{[0^*, \, 0^*]}(j)$ & $I_0^*$, $I_0^*$ & $D^2_4$ \\ \hline
$X_{[IV^*, \, 3, 1]}$ & $IV^*$ $I_3$ $I_1$ & $E_6A_2$ \\ \hline
$X_{[III^*, \, 2, 1]}$ & $III^*$ $I_2$ $I_1$ & $E_7A_1$ \\ \hline
$X_{[II^*, \, 1, 1]}$ & $II^*$ $I_1$ $I_1$ & $E_8$ \\ \hline
$X_{[4^*, \, 1, 1]}$ & $I_4^*$ $I_1$ $I_1$ & $D_8$ \\ \hline
$X_{[2^*, \, 2, 2]}$ & $I^*_2$ $I_2$ $I_2$ & $D_6A^2_1$ \\ \hline
$X_{[1^*, \, 4, 1]}$ & $I^*_1$ $I_4$ $I_1$ & $D_5A_3$ \\ \hline
$X_{[3, 3, 3, 3]}$ & $I_3$ $I_3$ $I_3$ $I_3$ & $A_2^4$ \\ \hline
$X_{[4, 4, 2, 2]}$ & $I_4$ $I_4$ $I_2$ $I_2$ & $A_3^2A_1^2$ \\ \hline
$X_{[5, 5, 1, 1]}$ & $I_5$ $I_5$ $I_1$ $I_1$ & $A^2_4$  \\ \hline
$X_{[6, 3, 2, 1]}$ & $I_6$ $I_3$ $I_2$ $I_1$ & $A_5A_2A_1$ \\ \hline
$X_{[8, 2, 1, 1]}$ & $I_8$ $I_2$ $I_1$ $I_1$ & $A_7A_1$ \\ \hline
$X_{[9, 1, 1, 1]}$ & $I_9$ $I_1$ $I_1$ $I_1$ & $A_8$ \\ \hline
\end{tabular}
\caption{\label{classificationRES in appB}Singular fiber types of extremal rational elliptic surfaces \cite{MP} are listed.}
\end{table}  
\endgroup

\end{document}